# High-Q optomechanical circuits made from polished nanocrystalline diamond thin films

S. Ummethala[1,+], P. Rath[1,+], G. Lewes-Malandrakis[2], D. Brink[2], C. Nebel[2], and W.H.P. Pernice[1,*]

[1]*Institute of Nanotechnology, Karlsruhe Institute of Technology, 76344 Eggenstein-Leopoldshafen, Germany*
[2]*Fraunhofer Institute for Applied Solid State Physics, Tullastr. 72, 79108 Freiburg, Germany*
+ The authors contributed equally to this work.

We demonstrate integrated optomechanical circuits with high mechanical quality factors prepared from nanocrystalline diamond thin films. Using chemomechanical polishing, the RMS surface roughness of as grown polycrystalline diamond films is reduced below 3 nm to allow for the fabrication of high-quality nanophotonic circuits. By integrating free-standing nanomechanical resonators into integrated optical devices, efficient read-out of the thermomechanical motion of diamond resonators is achieved with on-chip Mach-Zehnder interferometers. Mechanical quality factors up to 28,800 are measured for four-fold clamped optomechanical resonators coupled to the evanescent near-field of nanophotonic waveguides. Our platform holds promise for large-scale integration of optomechanical circuits for on-chip metrology and sensing applications.

[*]Email: wolfram.pernice@kit.edu

**Introduction**

Over the past decade, there has been a surge in applications of radiation pressure forces for optical tweezers [1-3], optomechanical cooling [4, 5] and optical transduction of mechanical motion [6, 7]. Optomechanical cooling by cavity feedback [8-10] recently allowed for preparing a macroscopic object in its quantum ground state [11], while active feedback cooling [12, 13] is actively pursued. Using optical methods for read out of mechanical vibrations provides unconstrained bandwidth and higher sensitivity compared to an electrical measurement, having enabled the observation of radiation pressure shot noise [14] and squeezing of light below the vacuum noise level [15]. In addition to purely optically transduced systems, optoelectromechanical systems using piezoelectric [16] and electrostatic [17-19] actuation combined with optical read-out are currently explored because of the benefit of stronger motion build-up.

A common requirement in optomechanical systems is the simultaneous realization of high optical measurement sensitivity and good performance of the mechanical resonator. In this respect materials that provide both good optical and mechanical properties are of special interest, making diamond a prime choice. Besides good optomechanical properties, diamond is highly interesting for fundamental research, due to the possibility of coupling to Nitrogen-Vacancy (NV) centers [20-23]. Because diamond is chemically inert and biocompatible, applications for sensing and detection of small compounds are on the horizon. Both research directions benefit from the large Young's modulus of 1100 GPa, which is one of the highest of all materials and allows for reaching high mechanical resonance frequencies. The high value of the Young's modulus of bulk diamond can be also achieved for thin films of polycrystalline diamond by carefully adjusting the deposition parameters [24]. Diamond mechanical resonators with high quality factors have since been successfully fabricated [25, 26]. Recently we have shown that mechanical resonators made from unpolished polycrystalline diamond with a doubly-clamped beam geometry can be integrated into photonic circuits [27]. Such circuits can be used for precision measurement of small displacements, forces and masses with the advantage of on-chip integration. However, the slot waveguide geometry used for such devices provides fabrication challenges when moving towards long mechanical resonators, for which high mechanical quality factors are expected.

Here, we report on the realization of extended diamond optomechanical devices. Using a resonator geometry suitable for the fabrication of long free-standing structures we obtain mechanical quality factors up to 28,800 at room temperature with high sensitivity. We fabricate four-fold clamped optomechanical resonators, so-called H-resonators [28], which are less susceptible to stiction forces and provide further flexibility for tailoring the mechanical resonance properties. By employing a chemomechanical polishing procedure, we reduce the as grown root mean square (RMS) surface roughness of polycrystalline diamond films below 3 nm RMS, measured on areas of 25 $\mu m^2$, enabling high-yield fabrication on a wafer-scale. Such optomechanical circuits are expected to provide lower propagation loss paired with higher mechanical quality factors for large-scale integration and advanced sensing applications.

**Diamond layer deposition and polishing**

In the current work, optomechanical circuits were realized on diamond-on-insulator (DOI) wafers. Such templates consist of a diamond thin film of several hundred nanometer thickness deposited on a buried oxide buffer layer, supported by a silicon substrate in analogy to silicon-on-insulator (SOI) wafers. Due to the high refractive index of diamond (2.4 at 1550 nm wavelength) the Diamond/$SiO_2$/Silicon stack offers good refractive index contrast for efficient waveguiding of near infrared (near-IR) light. The resulting tight optical confinement can thus be used for realizing high-quality optical resonators and interferometers [29-34]

Nanocrystalline diamond films were deposited on atomically flat silicon wafers which are thermally oxidized to a thickness of 2 µm. To initiate the diamond growth, firstly, a diamond nano-particle seed layer is deposited onto the $SiO_2$ film by ultrasonification for 30 minutes in a water based suspension of ultra-dispersed (0.1 wt %) nano-diamond particles of typically 5-10 nm size [35]. Then the samples are rinsed with deionized water and methanol. After dry blowing, the wafer is transferred into an ellipsoidal 2.54 GHz microwave plasma reactor [36] where diamond films with a thickness between 800 and 900 nm are grown at 3.5 kW microwave power, using 1 % $CH_4$ in 99 % $H_2$, at a pressure of 55 mbar and a temperature of 850 °C. Substrate rotation is applied to avoid angular non-uniformities arising from the gas flow. Growth rates are in the range of 1-2 µm/h. After growth, the samples are cleaned in concentrated $HNO_3$:$H_2SO_4$ to remove surface contaminants.

As grown nanocrystalline diamond films with a thickness of 600 nm typically have a RMS roughness of about 15 nm with a typical grain size on the order of 100 nm while isolated peaks can have a height up to several hundred nanometers. In order to facilitate fabrication of high resolution photonic structures with vertical side wall profile, the diamond films are polished using slurry based chemical mechanical planarization (CMP). This approach is commonly employed in the IC fabrication industry for the polishing of dielectric and metal interconnects, where a softer polyester based polishing pad is used with the aid of a colloidal silica at room temperature [37]. The technique does not require the use of expensive diamond grit, or cast iron scaifes. The CMP polishing was performed with a contact force of 120 N at a rotational frequency of 90 rpm and usage of 80 ml/min of polishing liquid containing silica particles. The polishing mechanism consists of the wet oxidation of the surfaces while the polishing fluid facilitates the attachment of silica particles to the diamond film. This is followed by shear removal of the particle due to forces from the polishing pad which is employed throughout. In our procedure, the experimental conditions closely follow the approach recently presented by Thomas et al. [38]. Using this CMP method, the thin film thickness variation of our diamond layers across the wafer surface can be polished without fear of film cracking.

After polishing, the average RMS roughness of the diamond layer is reduced from 15 nm down to 2.6 nm, on 25 µm$^2$ areas measured by a series of atomic force microscopy (AFM) scans. High resolution diamond nanophotonic circuits are fabricated using electron beam (e-beam) lithography on a JEOL 5300 50 kV system. We employ FoX-15 (HSQ 15%) negative tone resist with a thickness of 500 nm in order to provide sufficient protection against subsequent dry etching. The polished diamond surface now allows for fabrication of structures with smooth and straight sidewall profile. Fig. 10a and 10b show scanning electron microscopy (SEM) micrographs of developed HSQ resist after lithography, of the same device geometry on unpolished and polished diamond wafers, indicating the improved quality due to CMP. Comparing the patterns transferred from the resist half way into the diamond layer by reactive ion etching (RIE) (Fig. 10c and 10d) shows clearly that the resulting diamond structures have smoother surfaces and sidewalls. As roughness leads to scattering loss, integrated photonic circuits made from polished diamond are expected to exhibit less propagation loss.

In order to fabricate optomechanical circuits, a three step electron beam (e-beam) lithography process is employed. The initial lithography step using positive tone PMMA 950k 8% e-beam resist is carried out to define alignment markers for the subsequent lithography steps. Electron beam evaporation of a chromium-gold-chromium tri-layer (5nm-100nm-5nm) is then followed by lift-off in acetone resulting in metal alignment markers. The photonic and mechanical components are then patterned using HSQ. After development, the HSQ patterns are transferred into diamond by capacitively coupled RIE [39] with oxygen-argon plasma on an Oxford 80 Plasmalab system. In the final step, windows for releasing the mechanical resonators are patterned into PMMA. As PMMA does not sustain the diamond etch step in $O_2$/Ar chemistry, an additional chromium hard mask layer is used. After exposure, the PMMA pattern is transferred into the underlying chromium by wet etching. Then, the diamond in the unmasked regions is dry etched again with RIE to completely reveal the underlying oxide layer. The mechanical resonators are then released by isotropically etching the 2 µm thick $SiO_2$ in a buffered HF solution. Further details on the fabrication process can also be found in [27].

An optical micrograph of a fabricated nanophotonic device is shown in Fig. 11a. With the diamond-on-insulator platform we fabricate hundreds of devices on each chip, which allows us to parametrically scan the geometry parameters of on-chip components. Fabricated devices are characterized optically using transmission measurements, as described in our earlier work [27]. The nanophotonic circuits are equipped with a focusing grating coupler at the input port in order to allow light from external laser sources to be coupled into the on-chip waveguides [40]. Transmitted light is extracted out of the waveguide from a second grating coupler and is measured with a low noise photodetector (New Focus 2117).

In order to detect the thermomechanical vibration of the nanomechanical resonators, on-chip phase sensitive measurement is performed. For this purpose, Mach-Zehnder interferometers (MZI) are fabricated in proximity to the mechanical resonators as shown in the optical micrograph. The two arms of the MZI have a path difference of 100 µm to enable the interferometric phase detection. The output of each MZI is balanced due to symmetric fabrication of a mechanical resonator close to each of the interferometer arms. The etch depth of the grating couplers is critical in determining the bandwidth and coupling efficiency. With an etch depth of 45% on a 530 nm diamond film, a maximum transmission of 4.2% is obtained in the telecommunication C and L bands.

Mechanical resonators are designed in the shape of an 'H', as shown in the Fig. 11b. This design offers higher degrees of freedom to tune the resonance frequency by varying the length (*L*), width (*W*) and the size of the central section compared to a doubly clamped beam. In order to measure the thermomechanical response, the H-resonators are coupled to the waveguide of the MZI in close proximity via the evanescent field. For the fundamental in-plane oscillations of the H-resonator, the effective index of the mode travelling in the adjacent waveguide is altered and the corresponding phase shifts are translated into intensity variations by the MZI. Thus sensitive phase measurement can be obtained by operating on the quadrature point of the MZI transmission profile. A photonic crystal (PhC) section, as shown in the inset of Fig. 11b, is patterned across the center of the H-resonator in order to avoid power loss into the base plate. The PhC is designed with a lattice constant of a=600 nm and a hole radius of r=0.3a, which leads to a photonic bandgap centered at 1550 nm. Therefore the PhC provides an optical mirror which prevents light propagating along the waveguide to couple into the central rectangular area of the H-resonator.

**Results**

We simulate the mechanical modes of the H-resonators with finite element method (FEM) using COMSOL Multiphysics employing diamond's bulk Young's modulus as 1050 GPa, a density of 3.515 g/cm$^3$ and a Poisson ratio of 0.1. The simulated resonance frequencies of the fundamental in-plane mode as presented in Fig. 12a follow the trends predicted by the Euler-Bernoulli theory. The resonance frequencies measured in the experiment are plotted in Fig. 12b. The frequency is observed to scale linearly with the width *W* and as *1/L$^2$* with the length *L* of the H-resonator. The measured resonance frequencies are about 8% smaller than the simulated frequencies which can be attributed to differences in the Young's modulus of the nanocrystalline diamond deposited through CVD [24]. This means that the Young's modulus of our diamond layer is about five times larger than the one of silicon, which allows fabricating ultrahigh-frequency nano-optomechanical devices [41] more easily than with standard materials such as silicon-on-insulator.

A vacuum measurement setup is used to experimentally characterize the thermomechanical vibrations of the H-resonators. To reduce the air damping of mechanical resonators, the measurements are carried out at a pressure of 10$^{-7}$ mbar. Infrared light from a tunable laser is coupled into the input grating coupler of the MZI through a polarization controller. Light coming out of the device is detected by a low-noise photodetector. The rf signal caused by the thermomechanical motion is then analyzed, revealing the resonance frequencies of the mechanical resonators. The mechanical resonances presented here correspond to devices with gaps of 150 nm and 200 nm between the H-resonator and the evanescently coupling waveguide of the MZI arm. The mechanical Q factors of the H-resonators are obtained by fitting Lorentzian curves to the measured power spectral density. In a technique similar to the calibration of AFM cantilevers [42], the Brownian motion of the H-resonators can be used for calibration by comparing the measured spectral density to the calculated one. The power spectral density of the thermomechanical noise at resonance frequency is $S_y^{1/2} = \sqrt{4k_B TQ/m_{eff}(2\pi f)^3}$ where $k_B$ is the Boltzmann constant, $T$ the absolute temperature (300 K), $Q$ the mechanical quality factor, $m_{eff}$ the effective modal mass, and $f$ the resonance frequency. The effective modal mass of a particular mechanical mode can be calculated in the FEM simulation and is defined as $m_{eff} = \frac{m_0}{u_{max}^2 V} \iiint u(x,y,z)^2 \, dx \, dy \, dz$ where $V$ is the volume of the freestanding H-resonator, $m_0$ its physical mass, $u(x,y,z)$ the displacement and $u_{max}$ the maximum displacement. In this study, displacement sensitivities up to 14 fm/Hz$^{1/2}$ at an optical power of 3 mW on the device are obtained. We note that the use of higher optical powers for measurement further increase the displacement sensitivity due to an increase in the thermal energy of the system. A typical spectral density for the measured thermomechanical motion of one photonic device is shown in Fig. 13a. The two resonance peaks correspond to the two H-resonators in the arms of the MZI.

The quality factors extracted from thermomechanical measurements of devices of different lengths and widths are shown in Fig. 13b. For all devices we find high mechanical Q factors exceeding 10,000 at room temperature. For an H-resonator of 40 µm length and 800 nm width the best mechanical Q factor of 28,800 is obtained at 6.45 MHz, which is a factor of three higher than previously published quality factors for mechanical resonators in integrated optical circuits in diamond [27].

**Conclusion**
In summary, we have fabricated nanophotonic and optomechanical circuits from wafer-scale polycrystalline diamond thin films. By chemomechanical polishing, we reduced the as grown surface roughness by an order of magnitude below 3 nm RMS allowing for the realization of high-quality mechanical and optical components. Using mechanical resonators with high Q factors coupled to integrated Mach-Zehnder interferometers, we measured displacement sensitivities up to 14 fm/Hz$^{1/2}$. To the best of our knowledge, the mechanical Q of 28,800 obtained in this study is among the highest for mechanical structures of similar dimensions realized in polycrystalline diamond. This strongly emphasizes the use of polycrystalline films for the realization of photonic components without the need for sophisticated processing techniques as compared to single crystal diamond.

**Acknowledgements**
W.H.P. Pernice acknowledges support by the DFG grants PE 1832/1-1 & PE 1832/1-2 and the Helmholtz society through grant HIRG-0005. P. Rath and S. Ummethala acknowledge support by the Karlsruhe School of Optics & Photonics (KSOP). We also acknowledge support by the Deutsche Forschungsgemeinschaft (DFG) and the state of Baden-Württemberg through the DFG-Center for Functional Nanostructures (CFN) within subproject A6.4. The authors further wish to thank Silvia Diewald for assistance in device fabrication.

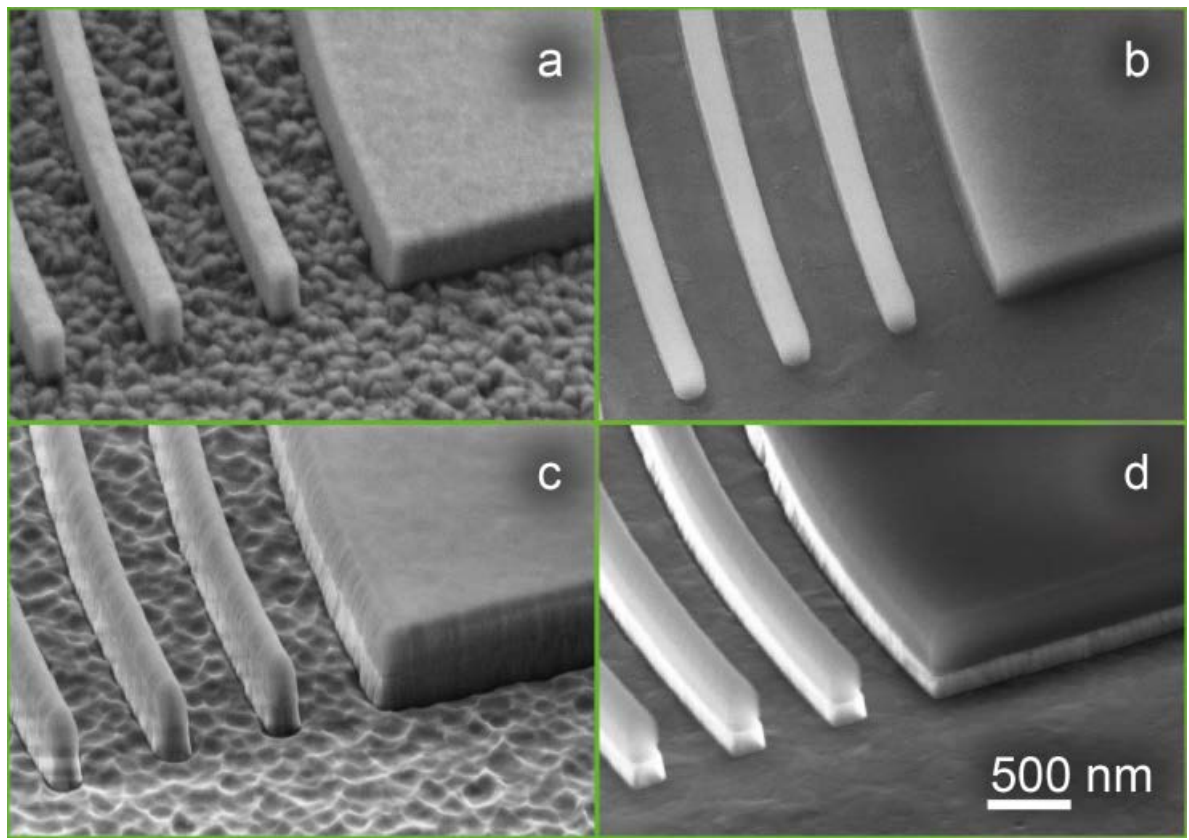

Figure 1: SEM images of fabricated nanophotonic components on unpolished (a,c) and polished (b,d) diamond thin films. Shown are grating coupler devices after electron-beam lithography and developing (a,b) and after pattern transfer by reactive ion etching (c,d). The HSQ ebeam resist remains on top of the structures after etching.

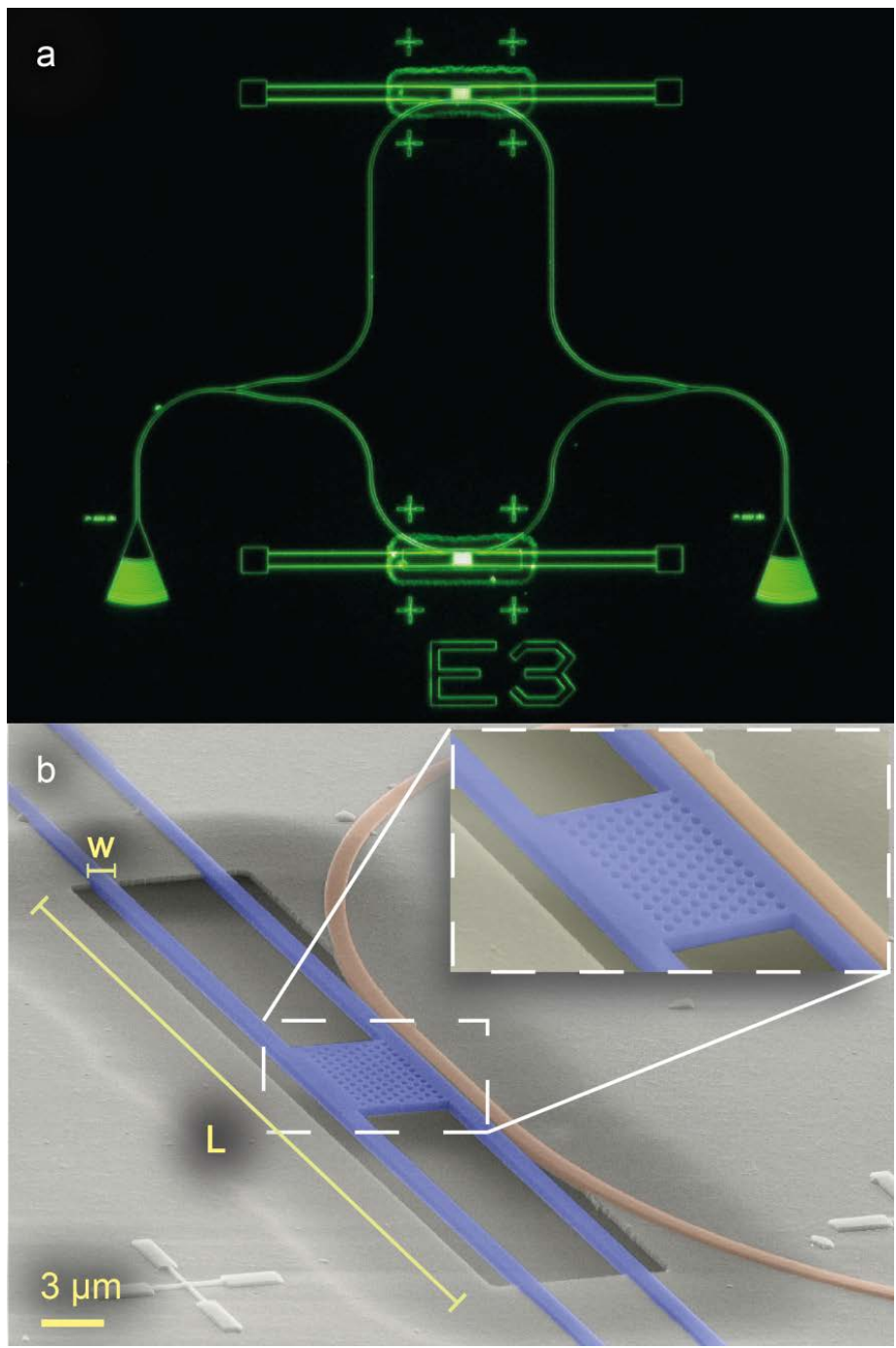

Figure 2: (a) Optical dark field micrograph showing one fabricated integrated optomechanical circuit, comprising grating coupler input/output ports, a Mach-Zehnder interferometer and two free-standing H-resonator devices in both the top and bottom arm. (b) False color SEM micrograph showing the released mechanical H-resonator. The inset shows the photonic crystal section. The length of the resonator is denoted by *L*, the width of the arms by *w*.

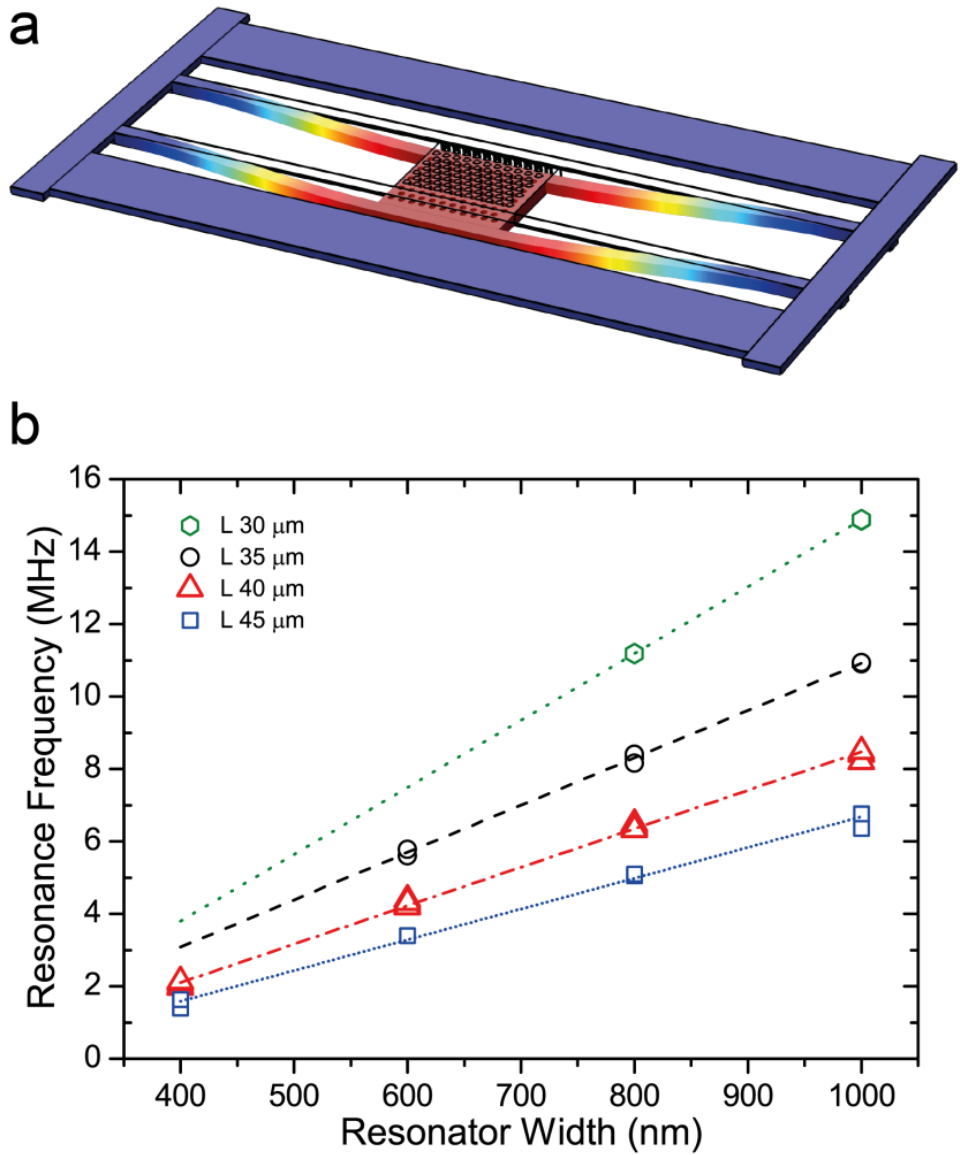

Figure 3: (a) 3D model FEM simulation of the fundamental in-plane mode of the H-resonator. (b) Measured dependence of the mechanical resonance frequency on the length $L$ and width $W$ of the arms. The resonance frequency increases linearly with increasing arm width and decreases as $1/L^2$ with increasing arm length. The dashed lines show the linear fits to the experimental data (markers).

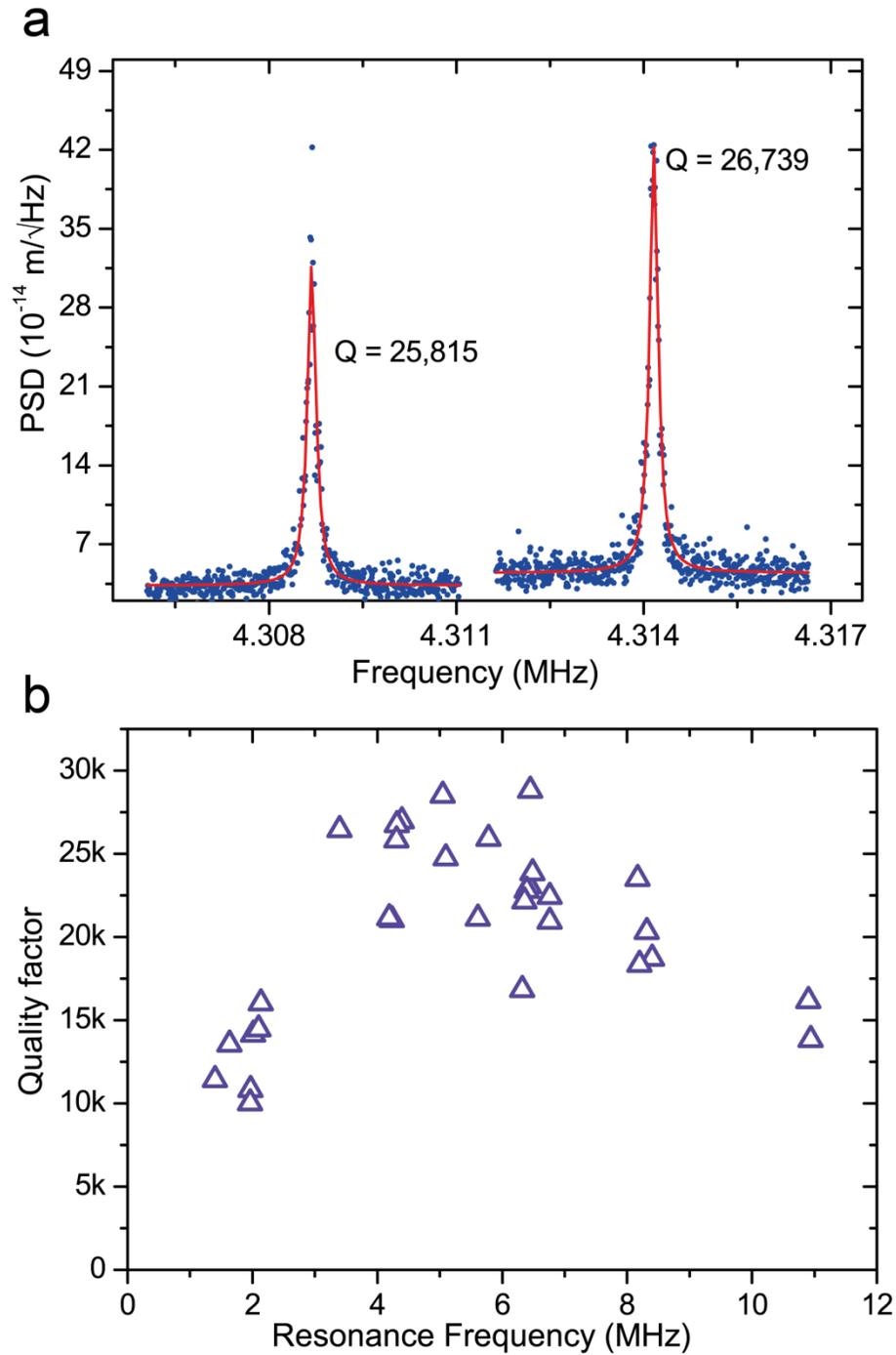

Figure 4: (a) The measured power spectral density (PSD) of a high-quality device showing two resonance peaks corresponding to the mechanical resonators in two arms of MZI. The mechanical quality factor is extracted from the Lorentzian fit to the data. (b) Measured mechanical Q-factors for resonators with different resonance frequencies. A maximum quality factor of 28,800 is found.